\documentclass[prd,preprint,preprintnumbers,nofootinbib,eqsecnum,superscriptaddress]{revtex4}

 \usepackage[dvips,final]{graphicx}
  \usepackage{amssymb}
   \usepackage{amsmath}
    \usepackage{amsfonts}
     \usepackage{epsfig}
      \usepackage{bm}

\usepackage{mathpazo}

\usepackage[section]{placeins}

\usepackage{multirow}
\usepackage{ctable}
\usepackage{booktabs}
\usepackage{array}
\usepackage{tabularx}
\usepackage{xcolor}
\usepackage{pstricks}

\begin{document}

\title{
Production of \bm{$\Lambda_c$} baryons at the LHC within \\
the \bm{$k_T$}-factorization approach \\ 
and independent parton fragmentation picture}

\author{Rafa{\l} Maciu{\l}a}
\email{rafal.maciula@ifj.edu.pl} \affiliation{Institute of Nuclear
Physics, Polish Academy of Sciences, Radzikowskiego 152, PL-31-342 Krak{\'o}w, Poland}

\author{Antoni Szczurek\footnote{also at University of Rzesz\'ow, PL-35-959 Rzesz\'ow, Poland}}
\email{antoni.szczurek@ifj.edu.pl} \affiliation{Institute of Nuclear
Physics, Polish Academy of Sciences, Radzikowskiego 152, PL-31-342 Krak{\'o}w, Poland}


\begin{abstract}
We calculate cross section for production of $D$ mesons and $\Lambda_c$ baryons 
in proton-proton collisions at the LHC. The cross section for 
production of $c \bar c$ pairs is calculated 
within $k_T$-factorization approach with the Kimber-Martin-Ryskin unintegrated gluon distributions obtained
on the basis of modern collinear gluon distribution functions.
We show that our approach well describes the $D^0$, $D^+$ and $D_s$
experimental data. We try to understand recent ALICE and LHCb data for $\Lambda_c$
production with the $c \to \Lambda_c$ independent 
parton fragmentation approach. 
The Peterson fragmentation functions are used. 
The $f_{c \to \Lambda_c}$ fragmentation fraction and 
$\varepsilon_{c}^{\Lambda}$ parameter for $c \to \Lambda_c$ are varied.
As a control plot we show transverse momentum distribution of different
species of $D$ mesons assuming standard values of the $f_{c \to D}$ 
fragmentation fractions known from the literature.
The fraction $f_{c \to \Lambda_c}$ neccessary to describe
the ALICE data is much larger than the average value obtained
from $e^+ e^-$ or $e p$ experiments.
No drastic modification of the shape of fragmentation function is allowed
by the new ALICE and LHCb data for $\Lambda_c$ production.
We also discuss a possible dependence of the
$\Lambda_c/ D^0$ baryon-to-meson ratio on rapidity and transverse momentum 
as seems observed recently by the ALICE and LHCb collaborations. 
Three different effects are considered: the value of
$\varepsilon_c^{\Lambda}$ parameter in Peterson fragmentation function
for $c \to \Lambda_c$, a kinematical
effect related to the hadronization prescription and a possible feed-down from
higher charmed-baryon excitations.
It seems very difficult, if not impossible, to understand
the ALICE data within the considered independent parton fragmentation scheme.
\end{abstract}


\maketitle

\section{Introduction}

Production of charm ($c \bar c$-pairs)
belongs in principle to the domain of perturbative physics.
The corresponding cross section can be calculated in
collinear-factorization approach. Leading-order (LO) calculation is
known to give too small cross section and rather next-to-leading order (NLO)
calculation must be performed (see \textit{e.g.} Refs.~~\cite{Nason:1989zy,Beenakker:1990maa}). An effective and efficient alternative is
$k_T$-factorization approach \cite{Catani:1990xk,Catani:1990eg,Collins:1991ty}. The $k_T$-factorization
provides a good description of $D$ meson production cross sections at 
RHIC \cite{Maciula:2015kea}, Tevatron \cite{Jung:2010ey} 
and at the LHC \cite{Maciula:2013wg,Karpishkov:2016hnx}. 

The production of $D$ mesons and/or nonphotonic leptons
requires a nonperturbative information about hadronization process.
To describe $D$ meson production fragmentation
functions (FFs) for $c \to D$ quark-to-meson transitions are usually 
included. In the context of heavy-flavour production the Peterson FFs \cite{Peterson:1982ak} are usually used. 
Also other scale-independent fragmentation functions can be found 
in the literature \cite{Kartvelishvili:1977pi,Collins:1984ms,Braaten:1994bz}.
The effects on $D$ meson production related to the different FFs models 
were discussed in Ref.~\cite{Maciula:2013wg}. 
Scale-dependent FFs for charm production 
that undergo evolution equations were proposed in Ref.~\cite{Kneesch:2007ey}.
The evolution equation leads by construction also to $g \to D$ fragmentation.
Such an approach gives a good description of the LHC data at large
transverse momenta ($p_T >$ 2-3 GeV) but overshoots experimental data
at low transverse momenta \cite{Kniehl:2009ar,Karpishkov:2016hnx}. 
However, the evolution approach was done only for massless quarks and 
it may be expected that inclusion of mass effect would probably change 
the results. Also it is not clear how initial conditions for evolution should be included.
A relatively large $g \to D$ transition leads to unusually large
$\sigma_{\mathrm{eff}}$, a parameter for double-parton scattering 
mechanism \cite{Maciula:2016wci}.
At large rapidities and low collision energies also subleading 
light quark/antiquark $q/\bar q \to D$ fragmentation may 
be important \cite{Maciula:2017wov}.

Recently the LHCb \cite{Aaij:2013mga} and very recently ALICE
\cite{Acharya:2017kfy} Collaborations obtained new results 
for $\Lambda_c$ production at the highest so far collision 
energy $\sqrt{s}$ = 7 TeV.
We wish to study whether the new LHCb and ALICE data can be described
consistently within the chosen scheme of 
calculation based on $c \to \Lambda_c$ fragmentation.
If yes, it would be interesting whether the $f_{c \to \Lambda_c}$
fragmentation fraction is consistent with those found in previous 
studies of $e^+ e^-$, $e p$ and $B$ meson decays.

\section{A sketch of the theoretical formalism}

\subsection{Parton-level calculations}

In the partonic part of our numerical calculations we follow the $k_{T}$-factorization approach. This approach is commonly known to be very efficient
not only for inclusive particle distributions but also for studies of kinematical correlations. It was also shown many times by different authors that it provides very good description of heavy quark production in proton-proton collisions at different energies. Some time ago it was successfully used for theoretical studies of $pp \to c \bar c \;\! X$ reaction at the LHC \cite{Maciula:2013wg,Karpishkov:2016hnx}. Very recently, this approach was also profitably applied \textit{e.g.} for $pp \to c \bar c + \mathrm{jet} \;\! X$ \cite{Maciula:2016kkx}, $pp \to c \bar c + \mathrm{2jets} \;\! X$ \cite{Maciula:2017egq} and $pp \to c \bar c c \bar c \;\! X$ \cite{Maciula:2013kd}.

According to this approach, the transverse momenta $k_{t}$'s (virtualities) of both partons entering the hard process are taken into account and the sum of transverse momenta of the final $c$ and $\bar c$ no longer cancels. Then the differential cross section at the tree-level for the $c \bar c$-pair production reads:
\begin{eqnarray}\label{LO_kt-factorization} 
\frac{d \sigma(p p \to c \bar c \, X)}{d y_1 d y_2 d^2p_{1,t} d^2p_{2,t}} &=&
\int \frac{d^2 k_{1,t}}{\pi} \frac{d^2 k_{2,t}}{\pi}
\frac{1}{16 \pi^2 (x_1 x_2 s)^2} \; \overline{ | {\cal M}^{\mathrm{off-shell}}_{g^* g^* \to c \bar c} |^2}
 \\  
&& \times  \; \delta^{2} \left( \vec{k}_{1,t} + \vec{k}_{2,t} 
                 - \vec{p}_{1,t} - \vec{p}_{2,t} \right) \;
{\cal F}_g(x_1,k_{1,t}^2) \; {\cal F}_g(x_2,k_{2,t}^2) \; \nonumber ,   
\end{eqnarray}
where ${\cal F}_g(x_1,k_{1,t}^2)$ and ${\cal F}_g(x_2,k_{2,t}^2)$
are the unintegrated gluon distribution functions (UGDFs) for both colliding hadrons and ${\cal M}^{\mathrm{off-shell}}_{g^* g^* \to c \bar c}$ is the off-shell matrix element for the hard subprocess. The extra integration is over transverse momenta of the initial
partons. We keep exact kinematics from the very beginning and additional hard dynamics coming from transverse momenta of incident partons. Explicit treatment of the transverse part of momenta makes the approach very efficient in studies of correlation observables. The two-dimensional Dirac delta function assures momentum conservation.
The unintegrated (transverse momentum dependent) gluon distributions must be evaluated at:
\begin{equation}
x_1 = \frac{m_{1,t}}{\sqrt{s}}\exp( y_1) 
     + \frac{m_{2,t}}{\sqrt{s}}\exp( y_2), \;\;\;\;\;\;
x_2 = \frac{m_{1,t}}{\sqrt{s}}\exp(-y_1) 
     + \frac{m_{2,t}}{\sqrt{s}}\exp(-y_2), \nonumber
\end{equation}
where $m_{i,t} = \sqrt{p_{i,t}^2 + m_c^2}$ is the quark/antiquark transverse mass. In the case of charm quark production at the LHC energies, especially in the forward rapidity region, one tests very small gluon longitudinal momentum fractions $x < 10^{-5}$.  

The matrix element squared for off-shell gluons is taken here in the analytic form proposed by Catani, Ciafaloni and Hautmann (CCH) \cite{Catani:1990eg}. It was also checked that the CCH expression is consistent with those presented later in Refs.~\cite{Collins:1991ty,Ball:2001pq} and in the limit of $k_{1,t}^2 \to 0$, $k_{2,t}^2 \to 0$ it converges to the on-shell formula.

The calculation of higher-order corrections in the $k_t$-factorization is much more complicated than in the case of collinear approximation.
However, the common statement is that actually in the $k_{t}$-factorization approach with tree-level off-shell matrix elements some part of real higher-order corrections is effectively included. This is due to possible emission of extra soft (and even hard) gluons encoded
in the unintegrated gluon densities. More details of the theoretical formalism adopted here can be found in Ref.~\cite{Maciula:2013wg}. 
  
In the numerical calculation below we have applied the Kimber-Martin-Ryskin (KMR) UGDF that is derived from a modified DGLAP-BFKL evolution equation \cite{Kimber:2001sc,Watt:2003mx} and has been found recently to work very well in the case of charm production at the LHC \cite{Maciula:2013wg}.
As discussed also in Ref.~\cite{Maciula:2016kkx} the $k_T$-factorization approach with the KMR UGDF gives results well consistent with collinear NLO approach.
For the calculation of the KMR distribution we used here up-to-date collinear MMHT2014 gluon PDFs \cite{Harland-Lang:2014zoa}.
The renormalization and factorization scales $\mu^2 = \mu_{R}^{2} =
\mu_{F}^{2} = \frac{m^{2}_{1,t} + m^{2}_{2,t}}{2}$ and charm quark mass 
$m_{c} = 1.5$ GeV are used in the present study. The uncertainties related to the choice of these paramteres were discussed in detail in Ref.~\cite{Maciula:2013wg} and will be not considered here.

\subsection{From quarks to hadrons}

Process of hadronization or parton fragmentation, i.e. transition from partons to hadrons, can be so far approached only through
phenomenological models. In principle, in the case of multi-particle final states the Lund string model \cite{Andersson:1983ia} and the cluster
fragmentation model \cite{Webber:1983if} are often used. However, the fragmentation of heavy quarks in the independent parton model (in non-Monte-Carlo calculations) is usually done with the help of fragmentation functions. These objects provide the probability for finding a hadron produced from a high energy quark or gluon.

According to the often used formalism, the inclusive distributions of charmed hadrons $h =D, \Lambda_c$ are obtained through a convolution of inclusive distributions of charm quarks/antiquarks and $c \to h$ fragmentation functions:
\begin{equation}
\frac{d \sigma(pp \rightarrow h X)}{d y_h d^2 p_{t,h}} \approx
\int_0^1 \frac{dz}{z^2} D_{c \to h}(z)
\frac{d \sigma(pp \rightarrow c X)}{d y_c d^2 p_{t,c}}
\Bigg\vert_{y_c = y_h \atop p_{t,c} = p_{t,h}/z} \;,
\label{Q_to_h}
\end{equation}
where $p_{t,c} = \frac{p_{t,h}}{z}$ and $z$ is the fraction of
longitudinal momentum of charm quark $c$ carried by a hadron $h =D, \Lambda_c$.
A typical approximation in this formalism assumes that $y_{c}$ is
unchanged in the fragmentation process, i.e. $y_h = y_c$. It was originally motivated for light hadrons
but is commonly accepted also in the case of heavy hadrons.

As a default set in all the following numerical calculations the
standard Peterson model of fragmentation function \cite{Peterson:1982ak}
with the parameters $\varepsilon_{c}^{D} = \varepsilon_{c}^{\Lambda} = 0.05$ is applied. The parameter will be varied only in the case of $c \to \Lambda_c$ transition. This choice of fragmentation function and parameters is based on our previous theoretical studies of open charm production at the LHC \cite{Maciula:2013wg}, where detailed analysis of uncertainties related to application of different models of FFs was done.

Another approach which makes use of phenomenological FFs is to assume that hadron is emitted in the same direction as charmed quark/antiquark, \textit{i.e.} $\theta_h = \theta_c$ that is equivalent to $\eta_h = \eta_c$, where $\eta_h$ and $\eta_c$ are hadron and quark pseudorapidities. We follow here the prescription presented in Ref.~\cite{Czech:2005vp} where the fragmentation quantity $z$ is defined by the equatioin $E_{h} = zE_{c}$.
 
Finally, the calculated cross sections for $D^{0}, D^{+}, D^{+}_{S}$ mesons and $\Lambda_c$ baryon should be normalized to the relevant fragmentation fractions.
For a nice review of the charm fragmentation fractions see Ref.~\cite{Lisovyi:2015uqa}.

\section{Results}

\subsection{Transverse momentum distributions of charmed
mesons and baryons}

\begin{figure}[!h]
\begin{minipage}{0.47\textwidth}
 \centerline{\includegraphics[width=1.0\textwidth]{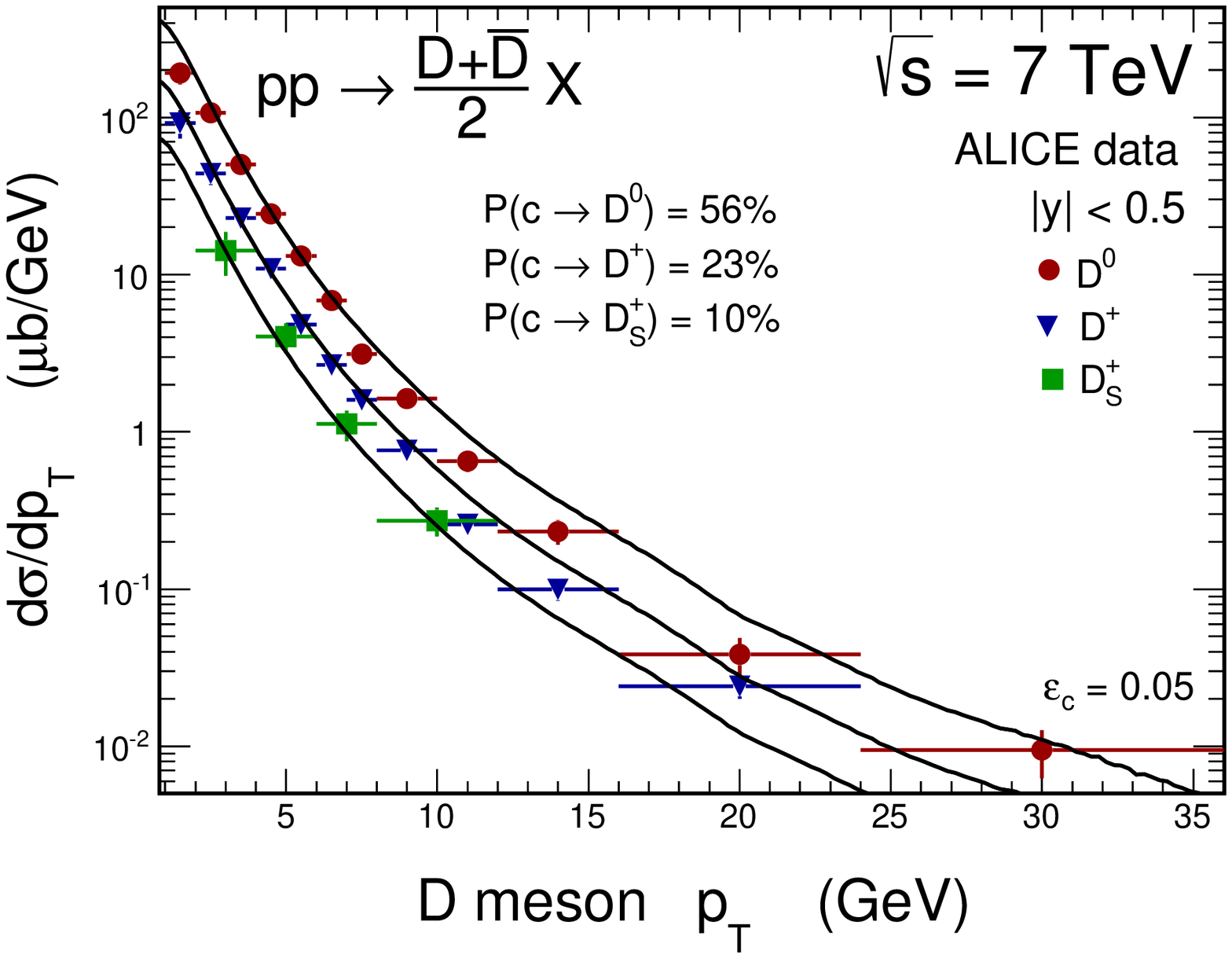}}
\end{minipage}
\begin{minipage}{0.47\textwidth}
 \centerline{\includegraphics[width=1.0\textwidth]{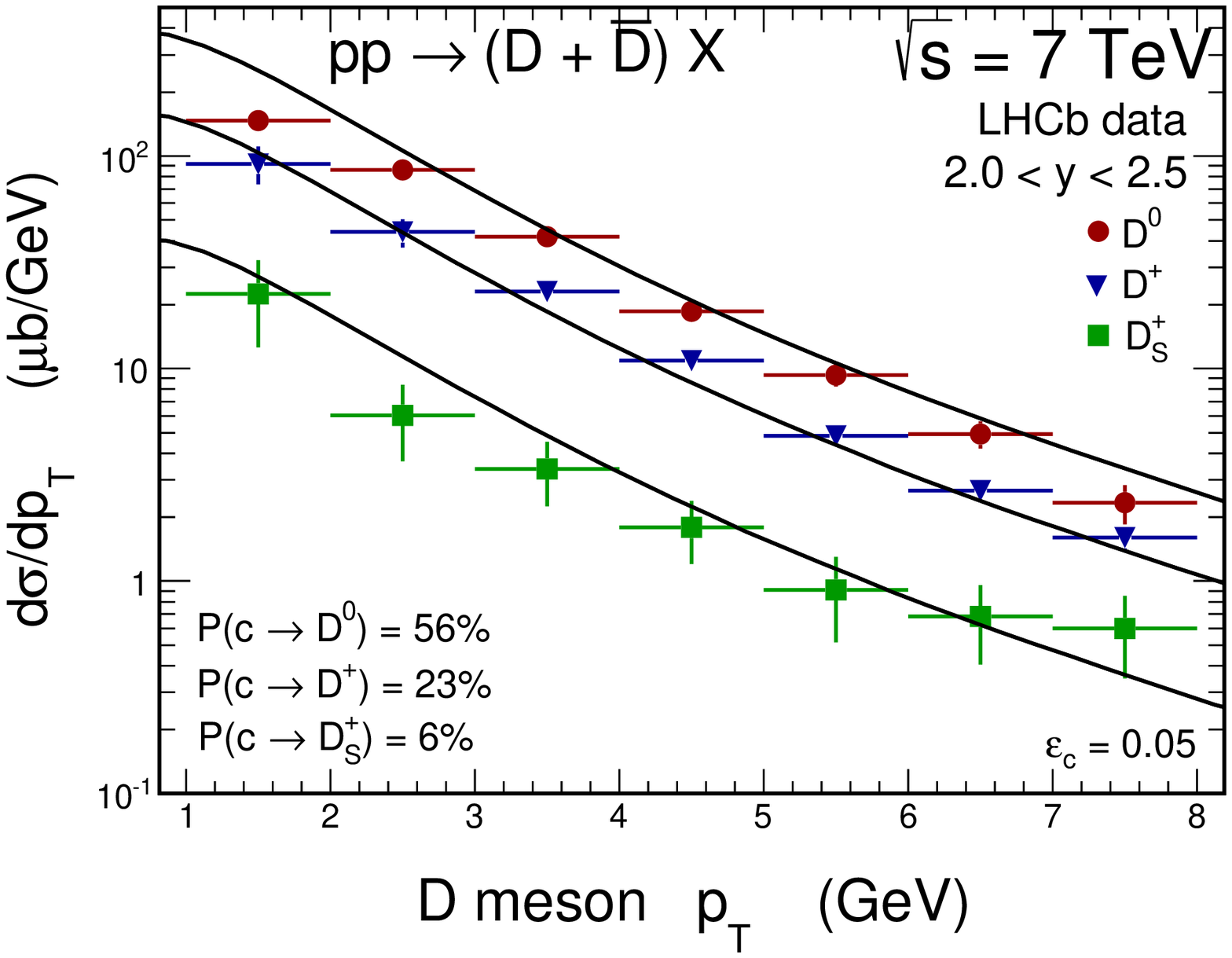}}
\end{minipage}
\caption{
\small Transverse momentum distribution of $D$ mesons for $\sqrt{s}$ = 7
TeV for ALICE (left panel) and LHCb (right panel). 
The experimental data points are taken from Refs.~\cite{Acharya:2017jgo} 
and \cite{Aaij:2013mga}, respectively.
}
 \label{fig:dsig_dpt_Dmesons}
\end{figure}

We start our presentation by showing results for $D$ meson
production. In Fig.~\ref{fig:dsig_dpt_Dmesons} we present transverse
momentum distributions of different open charm mesons - $D^0, D^+$, 
and $D_s$ for the ALICE (left panel) and the LHCb (right panel) kinematics.
Here, and throughout this subsection, the numerical results are obtained
within the standard fragmentation procedure with the assumption of
unchanged rapidity, \textit{i.e.} $y_{c} = y_{h}$, where $h=D,\Lambda_c$. In this calculation
we use standard Peterson fragmentation function with 
$\varepsilon_{c}^{D}= 0.05$ for $c \to D$ transition. 
The fragmentation fractions for charmed mesons are set to be 
$f_{c \to D^0}$ = 0.56 and $f_{c\to D^+}$ = 0.23 for both, ALICE 
and LHCb detector acceptance.
In the case of charmed-strange meson two different values of 
the fragmentation fraction are needed to fit both data sets with 
the same precision, \textit{i.e.} $f_{c \to D_S} = 0.06$ for LHCb 
and $0.10$ for ALICE. Both values of the fragmentation fraction for 
$c \to D_S$ transition are consistent with those extracted from 
combined analysis of charm-quark fragmentation fraction measurements 
in $e^+e^-$, $ep$, and $pp$ collisions \cite{Lisovyi:2015uqa}. 

\begin{figure}[!htbp]
\begin{minipage}{0.47\textwidth}
 \centerline{\includegraphics[width=1.0\textwidth]{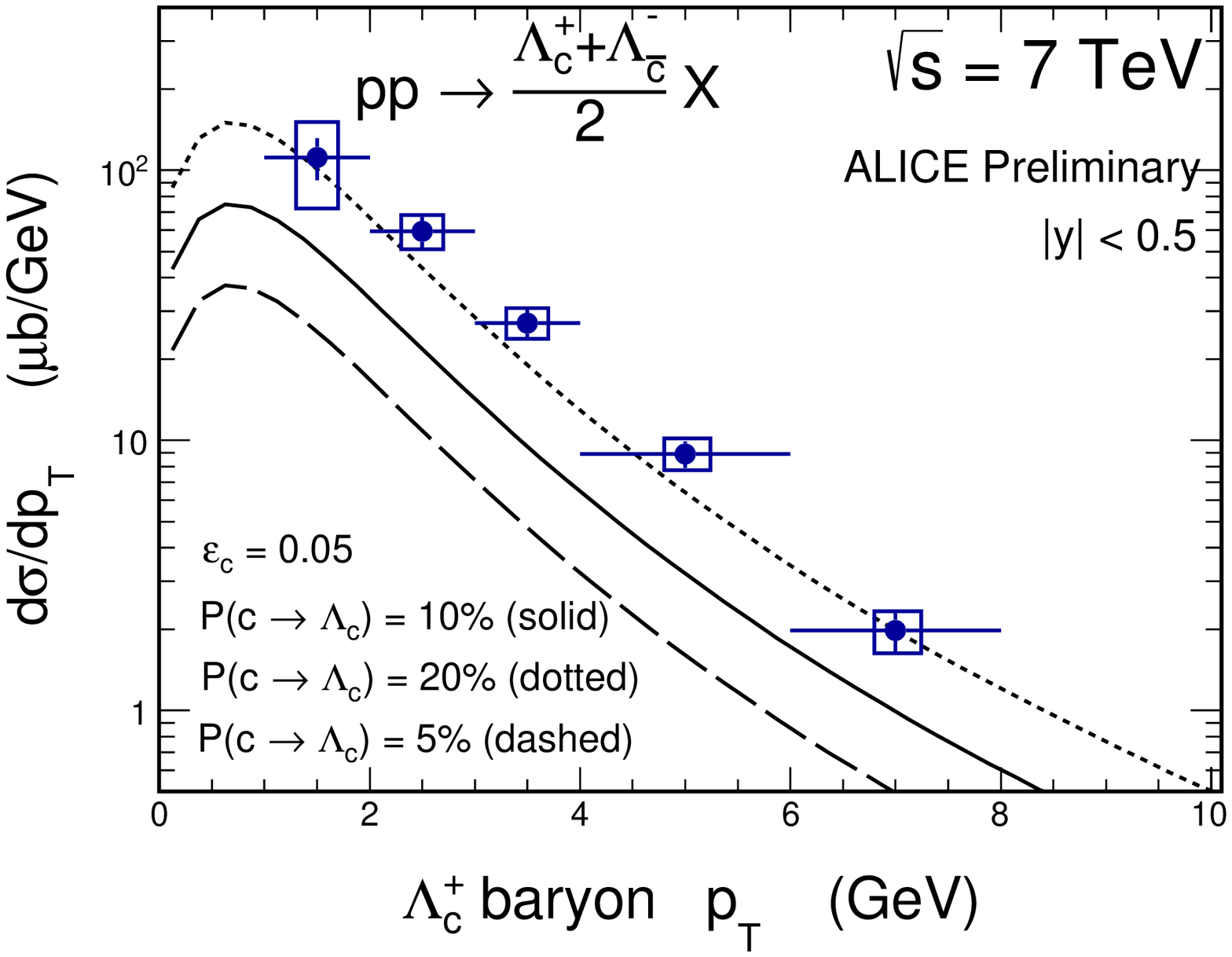}}
\end{minipage}
\begin{minipage}{0.47\textwidth}
 \centerline{\includegraphics[width=1.0\textwidth]{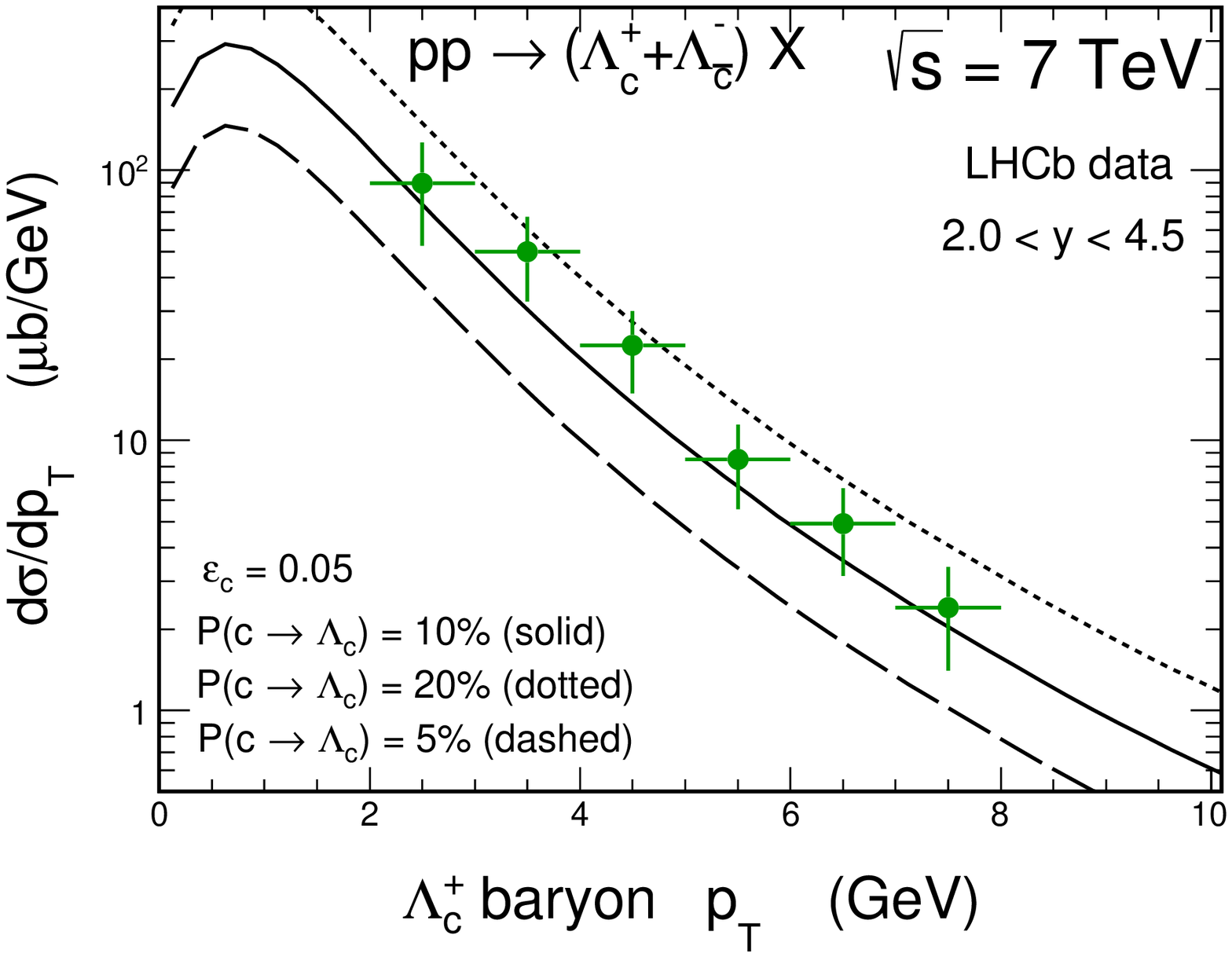}}
\end{minipage}
\caption{
\small Transverse momentum distribution of $\Lambda_c$ baryon 
for $\sqrt{s}$ = 7 TeV for ALICE (left panel) and LHCb (right panel).
The experimental data points are taken from Refs.~\cite{Acharya:2017kfy} 
and \cite{Aaij:2013mga}, respectively.
}
\label{fig:dsig_dpt_LambdaC}
\end{figure}

Having fixed all parameters of the theoretical approach in the context 
of open charm meson production we can proceed to the production 
of $\Lambda_c$ baryons. 
In Fig.~\ref{fig:dsig_dpt_LambdaC} we present transverse momentum
distribution of $\Lambda_c$ baryons for the ALICE (left panel) and 
the LHCb (right panel) kinematics. 
In this calculation we have also used the Peterson FF with the same
parameter $\varepsilon_{c}^{\Lambda} = 0.05$ (as a default) as 
for $c \to D$ transition. The three lines correspond to different values of 
$c \to \Lambda_c$ fragmentation fractions.
The dashed curve is for $f_{c \to \Lambda_c}$ = 0.05, as typical for
pre-LHC results. Clearly this result underpredicts both ALICE and LHCb
data. We show also result for increased fragmentation fractions, 
\textit{i.e.} $f_{c \to \Lambda_c}$ = 0.10 (solid line) and 0.15 (dashed line).
The agreement between data and the theory predictions with the increased
$f_{c \to \Lambda_c}$ becomes better. However, a visible difference
appears in the observed agreement for the mid-rapidity ALICE and 
for forward LHCb regimes. Taking $f_{c \to \Lambda_c}$ = 0.10 
we are able to describe the LHCb data quite well 
but we still underestimate the ALICE data by a factor $\sim 2$ in 
the whole considered range of transverse momenta. The shapes of the
transverse momentum distributions are well reproduced in both ALICE and LHCb cases.     
In order to get right normalization in the case of the ALICE measurement
we need to take $f_{c \to \Lambda_c}$ = 0.20 which is much bigger than
the numbers found in previous studies (see \textit{e.g.} a review 
in Ref.~\cite{Lisovyi:2015uqa}).

\begin{figure}[!htbp]
\begin{minipage}{0.47\textwidth}
 \centerline{\includegraphics[width=1.0\textwidth]{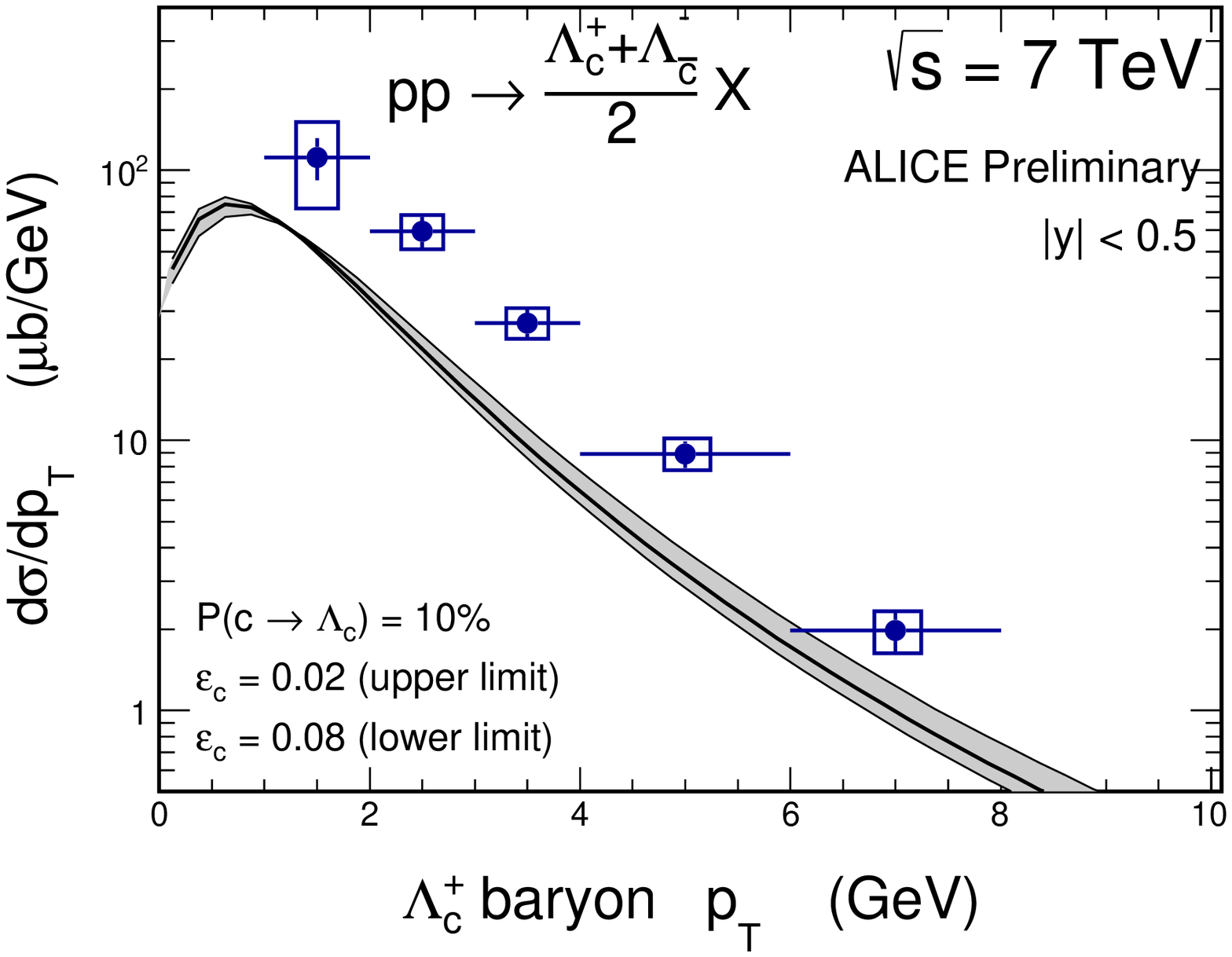}}
\end{minipage}
\begin{minipage}{0.47\textwidth}
 \centerline{\includegraphics[width=1.0\textwidth]{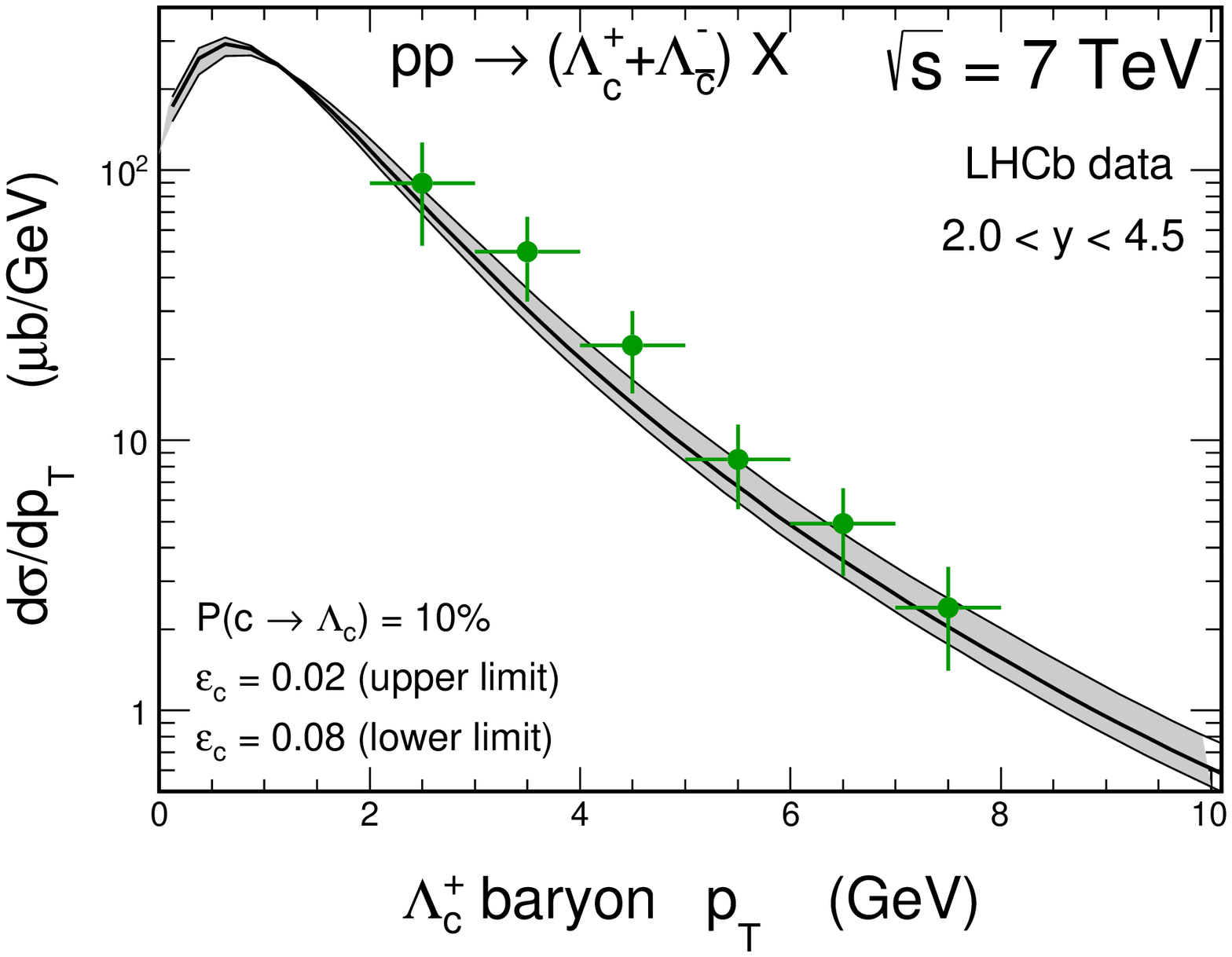}}
\end{minipage}
\caption{
\small Transverse momentum distribution of $\Lambda_c$ baryon 
for $\sqrt{s}$ = 7 TeV for ALICE (left panel) and LHCb (right panel)
for differrent values of the $\varepsilon_{c}^{\Lambda}$ parameter in the Peterson
fragmentation function for the $c \to \Lambda_c$ transition.
}
 \label{fig:dsig_dpt_Lambdac_epsilon}
\end{figure}
 
Can the situation change when the shape of the $c \to \Lambda_c$ 
fragmentation function is different?
As an illustration in Fig.~\ref{fig:dsig_dpt_Lambdac_epsilon}
we show results for different values of the 
$\varepsilon_{c}^{\Lambda}$ parameter in the Peterson fragmentation
function. As can be seen from the figure a drastic modification of 
the parameter changes the shape of the distribution in 
a moderate way. Here, the LHCb data suggests a choice of rather harder 
FF, \textit{i.e.} smaller $\varepsilon_{c}$ parameter, than in 
the case of the $c \to D$ transition. 
Going to the region of very small transverse momenta $p_{T}
\lesssim m_{c}$ the upper and lower limits start to reverse. However, 
it seems that playing with the shape of the fragmentation function will 
not help to understand the huge enhancement of $\Lambda_c$ production
observed by the ALICE Collaboration.
 
\subsection{Possible reasons for dependence of 
the \bm{$\Lambda_c/D^0$} ratio on transverse momentum and rapidity}

The ALICE Collaboration also reported much larger $\Lambda_c/D^0$ baryon-to-meson ratio
\cite{Acharya:2017kfy} than measured by the LHCb Collaboration
\cite{Aaij:2013mga}. Is this real effect? This observation seems to be
consistent with the conclusions drawn above. Here we want to discuss 
possible reasons for the transverse momentum and
dependence of the ratio.

\begin{figure}[!htbp]
\begin{minipage}{0.47\textwidth}
 \centerline{\includegraphics[width=1.0\textwidth]{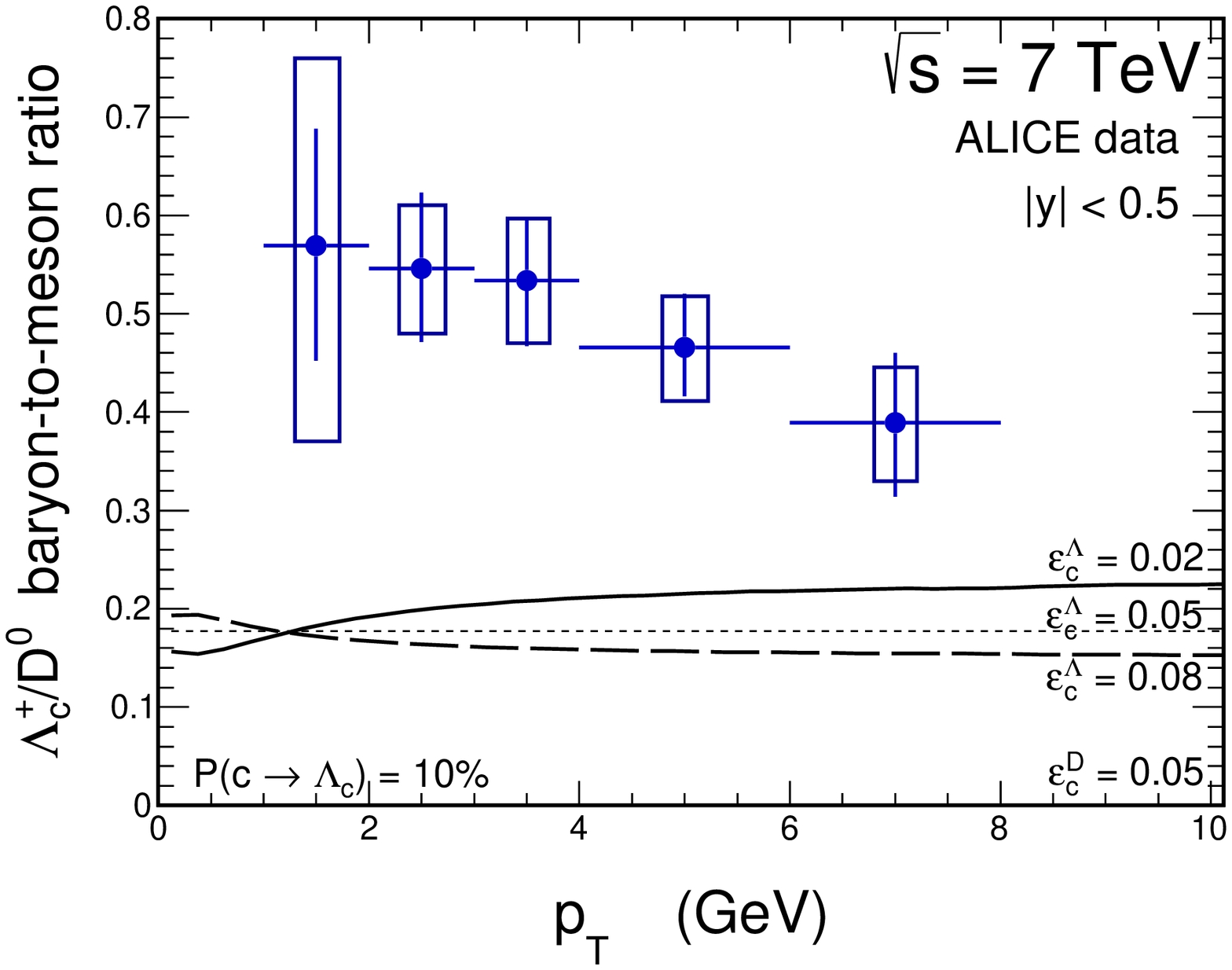}}
\end{minipage}
\begin{minipage}{0.47\textwidth}
 \centerline{\includegraphics[width=1.0\textwidth]{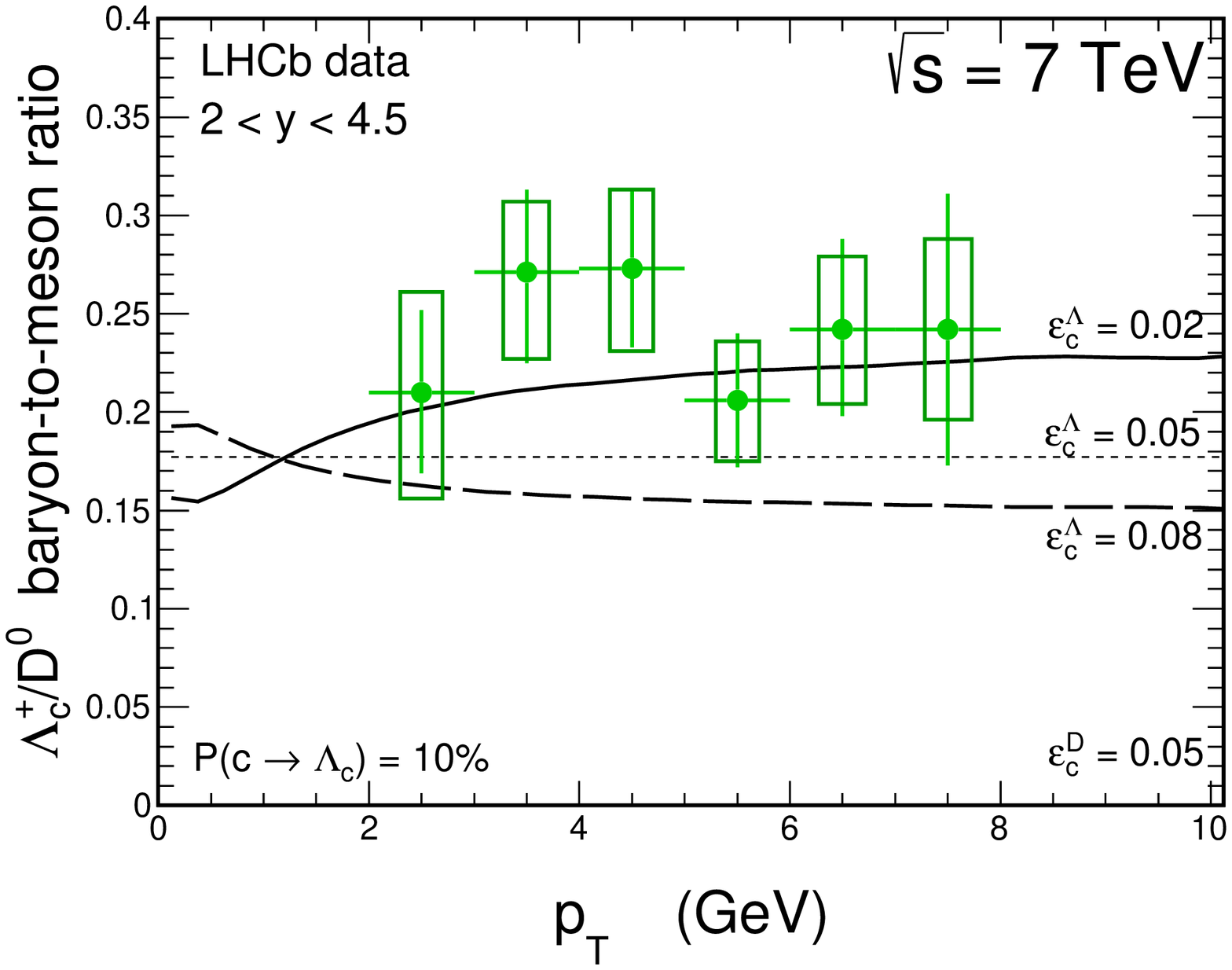}}
\end{minipage}
\caption{
\small Transverse momentum dependence of the
$\Lambda_c/D^0$ baryon-to-meson ratio for ALICE (left) and LHCb (right)
for different choices of the $\varepsilon_{c}^{\Lambda}$
parameter for $c \to \Lambda_c$ transition in the Peterson 
fragmentation function.
}
 \label{fig:ratio_pt_different_epsilons}
\end{figure}

The fragmentation function for $c \to \Lambda_c$ does not need to be the
same as for the $c \to D$ transition. This can be included even using 
the Peterson FFs by choosing different $\varepsilon_c$ parameter 
for fragmentation to $D$ meson and to $\Lambda_c$ baryon.
In Fig.~\ref{fig:ratio_pt_different_epsilons} we show the ratio
$\Lambda_c/D^0$ as a function of transverse momentum for ALICE 
(left panel) and LHCb (right panel). 
Here we keep the $\varepsilon_{c}^{D} = 0.05$ for the $c \to D$
fragmentation and take three different values 
$\varepsilon_{c}^{\Lambda} = 0.02$
(solid line), $0.05$ (dotted line), and $0.08$ (dashed line) for the $c \to
\Lambda_c$ transition. Again, taking smaller $\varepsilon_c$ parameter
for $c \to \Lambda_c$ case than for $c \to D$ fragmentation we are able 
to stay in touch with the LHCb experimental data, however, we 
visibly underestimate the ALICE data. A similar situation was reported 
in Ref.~\cite{Acharya:2017kfy} where the ALICE measurements were 
compared with results of different Monte Carlo event generators.
  
In the literature the fragmentation of $c \to h$, where $h$ is charmed
meson or baryon, is usually done assuming that $y_{h} = y_c$. 
This is an approximation which was, in fact, not discussed carefully
in the literature and which was originally derived for massless particles (both, parton and hadron).
Here we wish to discuss shortly how
the situation would change when different fragmentation scheme is used. To illustrate the issue we shall consider another approximation:
$\eta_{h} = \eta_c$, \textit{i.e.} we assume that the charmed hadron is
emitted in the same direction (in the proton-proton center of mass) 
as the quark/antiquark (see also Ref.~\cite{Czech:2005vp}).
Within this approach implicitly masses of incident parton and final hadron are taken into account.  

\begin{figure}[!htbp]
\begin{minipage}{0.47\textwidth}
 \centerline{\includegraphics[width=1.0\textwidth]{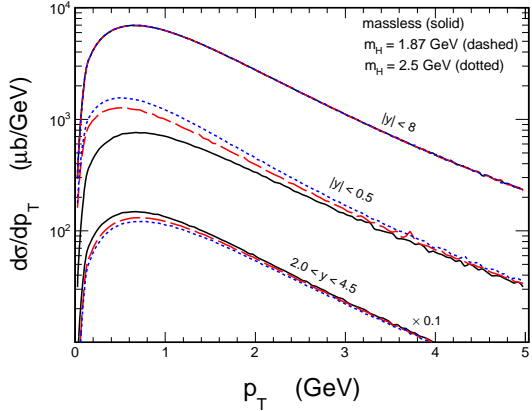}}
\end{minipage}
\caption{
\small Transverse momentum dependence of the cross section for
different intervals of rapidities and for different approaches to fragmentation procedure.
The dotted and dashed lines correspond to the $\eta_{h} = \eta_c$ prescription
for fragmentation. The solid lines are calculated with the standard $y_{h} = y_c$ approximation.
}
 \label{fig:dsig_dpt_different_frag}
\end{figure}

Let us try to compare results of such two approximations in Fig.~\ref{fig:dsig_dpt_different_frag}.
The solid lines correspond to the standard $y_{h} = y_c$ approximation, while the dashed and dotted lines are calculated with
the $\eta_{h} = \eta_c$ prescription, with $m_{h} = 1.87$ and $2.5$ GeV, respectively.
Here we have taken $m_{h} = 2.5$ GeV which corresponds on average to the masses of $\Sigma_c$ baryons \cite{PDG}.
We observe that the latter approach leads to
an enhancement of the cross section for small transverse momenta at midrapidities, which is the region relevant for the ALICE experiment. 
Simultaneously, an opposite effect is observed in the forward rapidity region, where the cross section is slightly lowered with respect to the standard
"massless" prescription.       

\begin{figure}[!htbp]
\begin{minipage}{0.47\textwidth}
 \centerline{\includegraphics[width=1.0\textwidth]{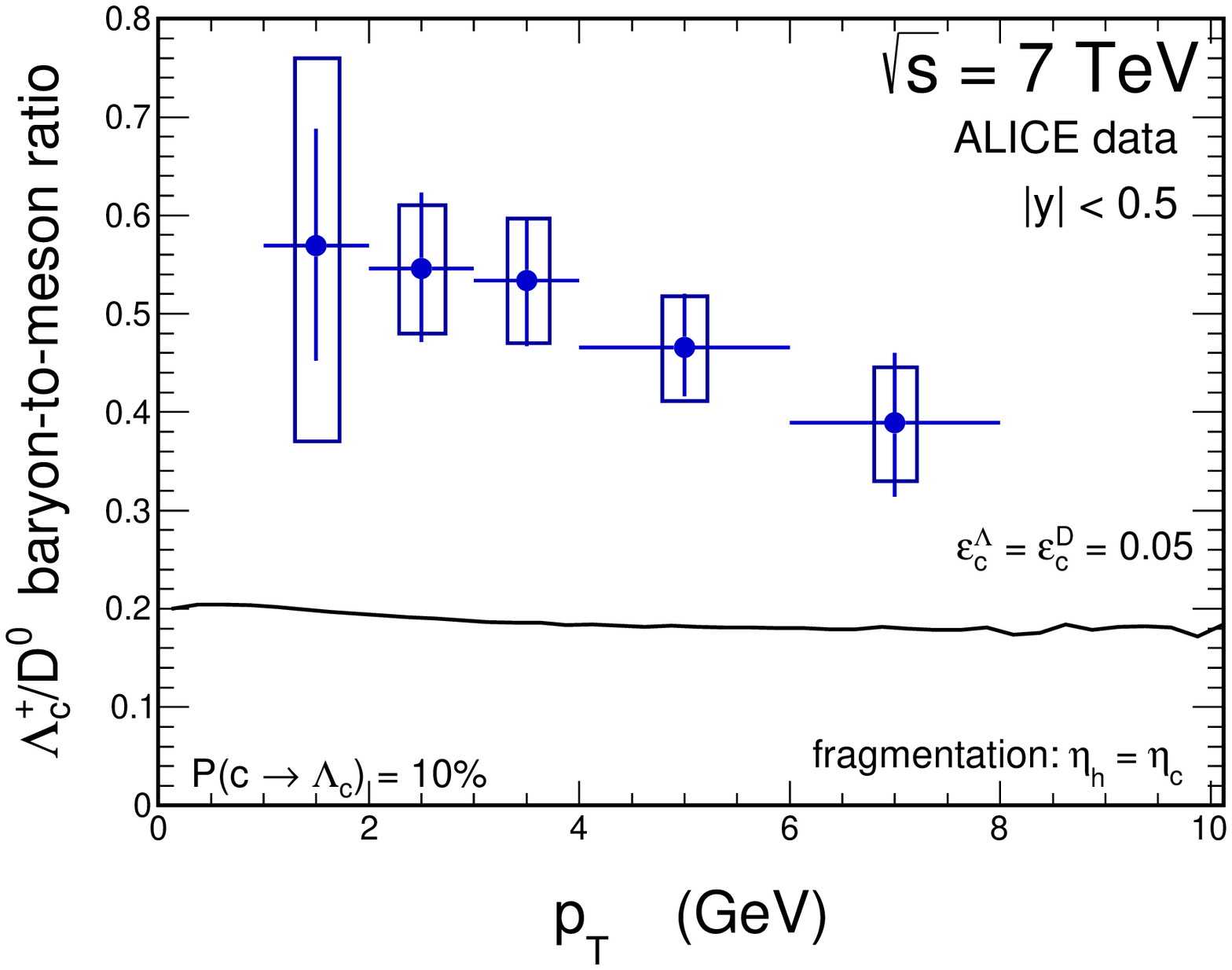}}
\end{minipage}
\begin{minipage}{0.47\textwidth}
 \centerline{\includegraphics[width=1.0\textwidth]{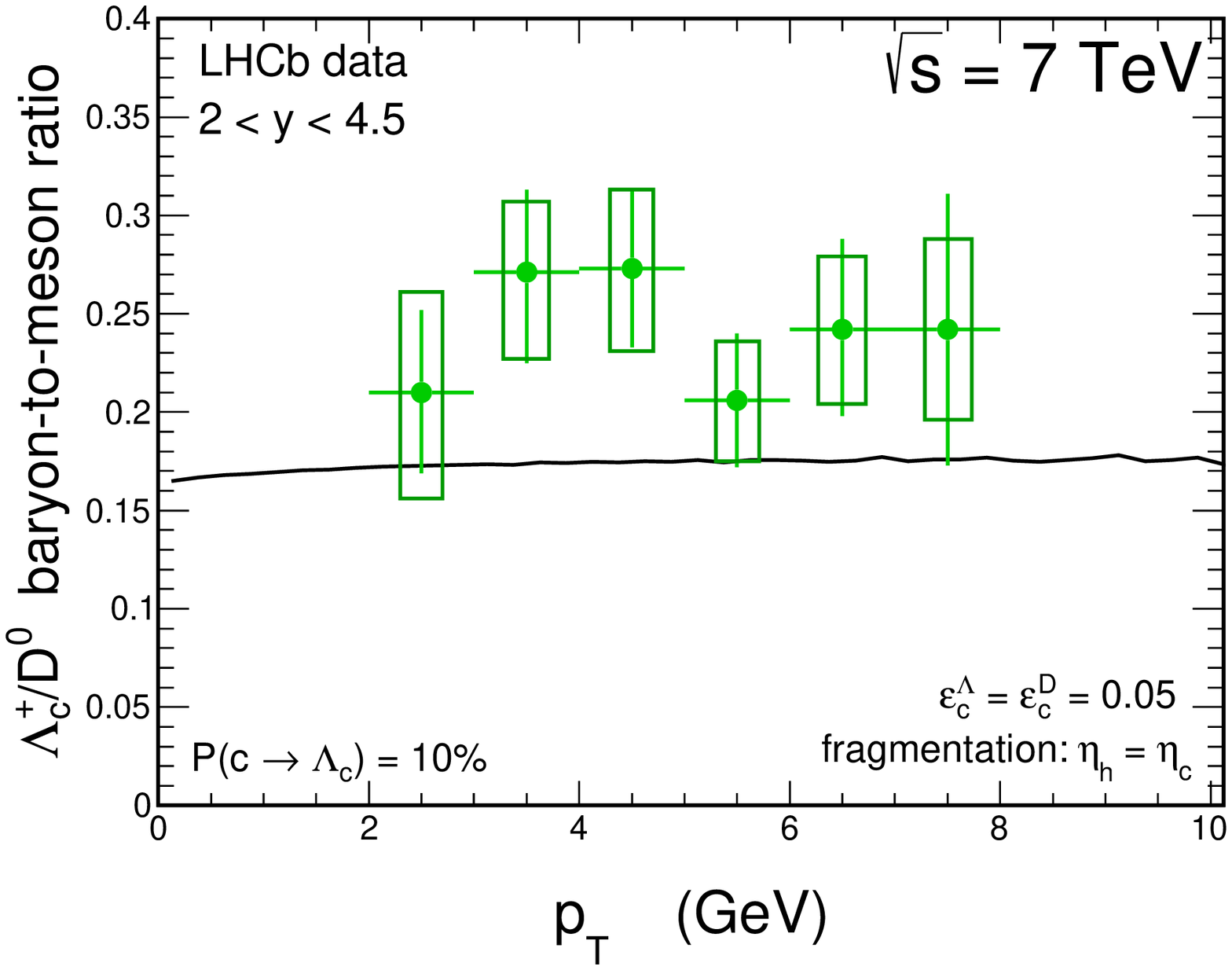}}
\end{minipage}
\caption{
\small Transverse momentum dependence of the $\Lambda_c/D^0$ 
baryon-to-meson ratio for ALICE (left) and LHCb (right)
for the $\eta_{h} = \eta_c$ approximation. Here only one (default) 
set of $\varepsilon_{c}$ parameters for the Peterson fragmentation functions
was used.
}
 \label{fig:ratio_fixedeta}
\end{figure}

Now we wish to show how big is the effect related to the new fragmentation scheme on the $\Lambda_c/D^0$ 
baryon-to-meson ratio. In Fig.~\ref{fig:ratio_fixedeta} we observe only a small dependence of the ratio on the meson/baryon transverse momentum.
However, comparing the left (ALICE) and right (LHCb) panels we
observe a slightly different value of the ratio. The change of 
the approximation leads to an enhancement of the ratio for the ALICE
kinematics compared to the standard approch used in the literature.
The enhancement is, however, only of the order of 10 \%, but in 
the right direction. Here we wanted only to illustrate the effect
of the enhanced production for ALICE related to the approximation
made for fragmentation, so the $P(c \to \Lambda_c)$ is kept
constant at the value known from other processes.
The observed enhancement seems, however, too small to explain
the gigantic enhancement observed by the ALICE collaboration.

So far only $\Lambda_c$ baryons were measured in proton-proton
scattering. However, there is a multitude of prediced and/or observed
singly charmed baryons: 6 of spin 1/2 and 6 of spin 3/2.
We have no idea about their production in proton-proton collisions.
Some of them could lead to a feed down to $\Lambda_c$ baryon \cite{PDG}.
Some of them decay weakly and could, in principle, be eliminated.
Examples of interest are $\Sigma_c$ baryons both for $J=$ 1/2
($\Sigma_c$(2455)) and $J=$ 3/2 ($\Sigma_c$(2520)).
The $\Sigma_c$ baryons are known to decay almost in 100 \% into 
$\Lambda_c$ and a pion \cite{PDG}.

\begin{figure}[!htbp]
\begin{minipage}{0.47\textwidth}
 \centerline{\includegraphics[width=1.0\textwidth]{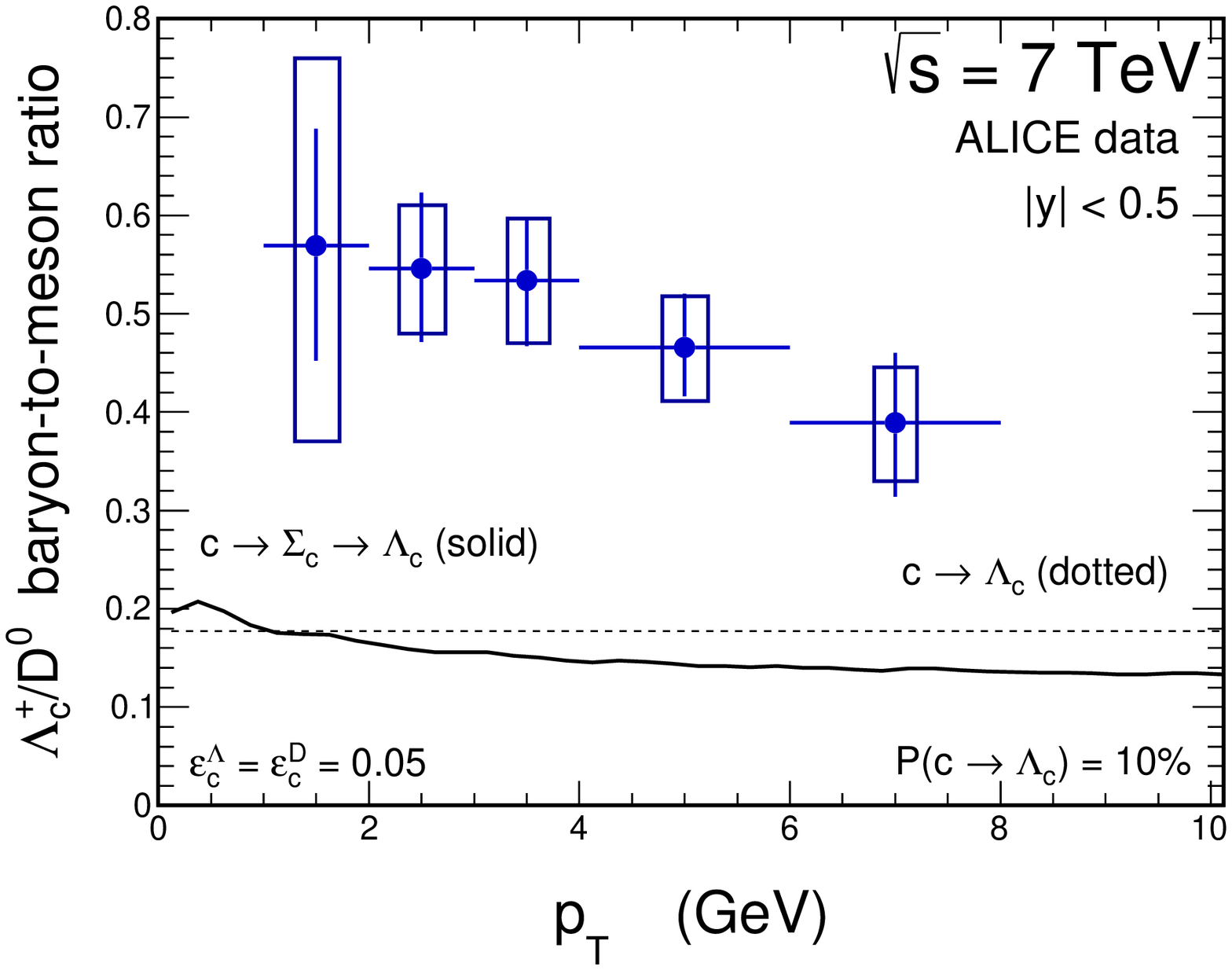}}
\end{minipage}
\begin{minipage}{0.47\textwidth}
 \centerline{\includegraphics[width=1.0\textwidth]{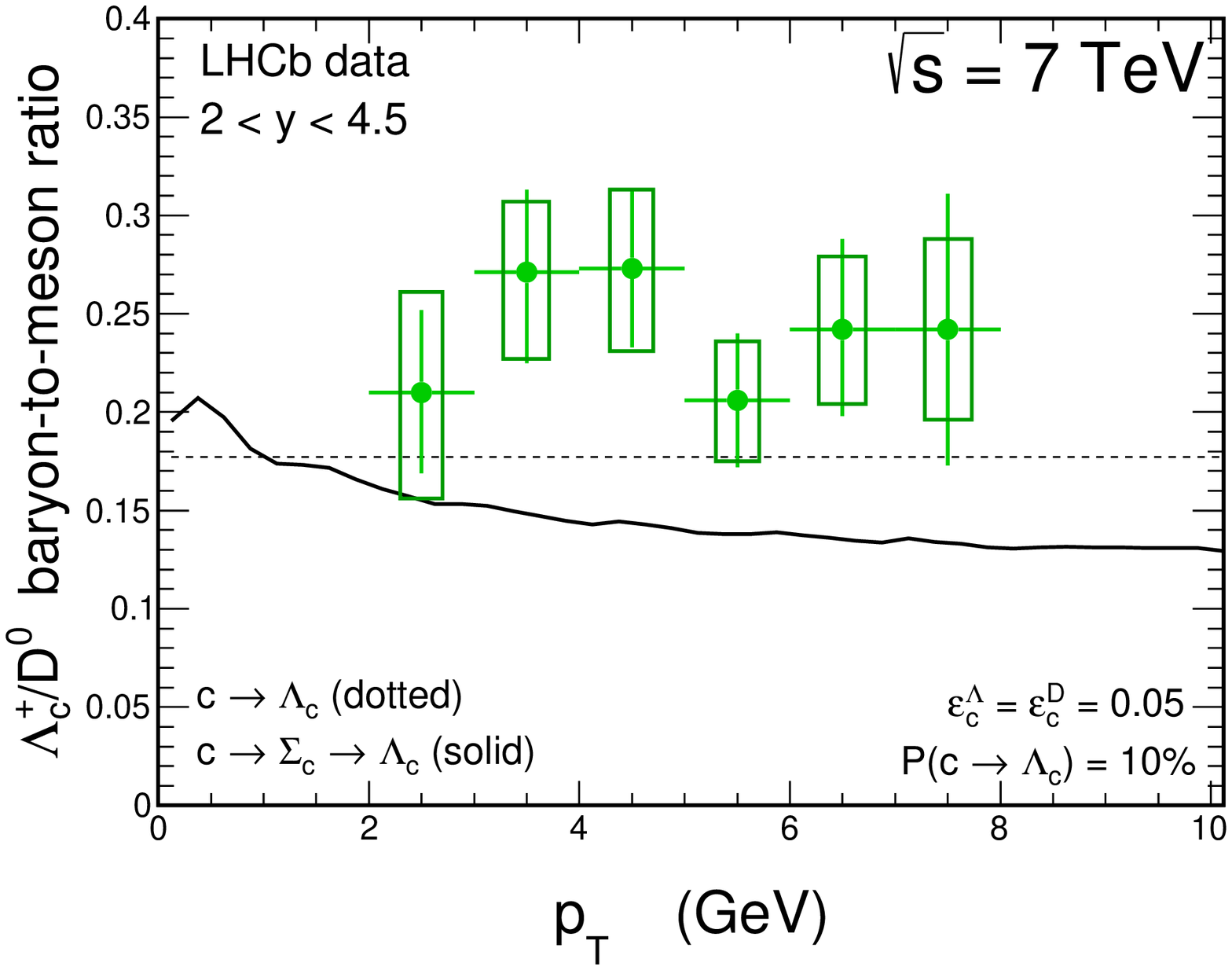}}
\end{minipage}
\caption{
\small Transverse momentum dependence of the $\Lambda_c/D^0$ 
baryon-to-meson ratio for ALICE (left) and LHCb (right)
for the feed-down mechanism (solid lines). Here the standard $y_h = y_c$ fragmentation procedure with only one (default) 
set of $\varepsilon_{c}$ parameter for the Peterson fragmentation function
was used. The dashed lines shown for reference, correspond to direct production $c \to \Lambda_c$.
}
 \label{fig:dsig_dpt_sequential_decay}
\end{figure}

So far we have implicitly assumed only a direct production of $\Lambda_c$
baryons (as represented by a Peterson fragmentation function with an 
a priori unknown $\varepsilon$ parameter).
In principle, $\Lambda_c$ baryons do not need to be produced exclusively
directly but may come from a feed down mechanism from excited baryonic states. 
The feed-down mechanism could modify transverse momentum distributions.
The decay is done in a small Monte Carlo code. We assume isotropic decay
in the rest frame of the excited state. Then Lorentz boosts are
performed to get distribution in the laboratory 
(proton-proton center-of-mass) frame. 
In Fig.~\ref{fig:dsig_dpt_sequential_decay} we show what can be the
effect of such a feed down. This mechanism leads to a small enhancement of the $\Lambda_c/D^0$ ratio
at small transverse momenta but it also causes its lowering at larger $p_{T}$'s.
We conclude that the feed down mechanism cannot explain
the enhanced production observed by the ALICE Collaboration.

\section{Conclusions}

We have discussed production of $\Lambda_c$ baryon in proton-proton
collisions at the LHC energy.
The cross section for $c \bar c$ production has been calculated
within the $k_t$-factorization approach. The Kimber-Martin-Ryskin
unintegrated gluon distribution has been used.
This combination assures realistic charm rapidity and transverse
momentum distributions.
The hadronization has been performed within a simple independent parton fragmentation 
function formalism. Peterson fragmentation
function has been used in the present analysis.
Varying the value of the $\epsilon_c$ parameter for the $c \to \Lambda_c$ tranistion  modifies the shape
of transferse momentum distribution. A good agreement with
the data (shape only) is obtained with $\epsilon_c$ parameter of similar size as that
found for $D$ mesons.

We find that the fragmentation fraction 
$f_{c \to \Lambda_c}$ = 0.1 - 0.15 describes the recent data
of the LHCb collaboration but fails to describe the new ALICE data.
Even for LHCb this number is slightly bigger than the values from 
the compilation of world results \cite{Lisovyi:2015uqa} obtained 
from experimental data on 
$e^+ e^-$ and $e p$ and $B$ meson decays.

The interpretation of the increased fragmentation fraction
$c \to \Lambda_c$ is at present not clear and requires further studies, 
both on the theoretical and experimental side.

We have also addressed the issue of possible dependence of the 
$\Lambda_c / D^0$ ratio on rapidity and transverse momentum. Three different effects have
been studied. We have discussed how much the effect may depend on the not
well known $\epsilon_c$ parameter in the Peterson fragmentation function
for the $c \to \Lambda_c$ fragmentation. Only a small effect has been found.
In addition, we have shown that a different treatment of the 
$c$ quark/antiquark fragmentation may slightly enhance the
production of $\Lambda_c$ with respect to $D$ mesons at midrapidities.
This effect is of purely kinematical origin and should not be visible in
pseudorapidity distributions.
Finally we have discussed whether indirect production of $\Lambda_c$
baryons could be related to the recent ALICE observation.
For example we have considered possible feed-down from $\Sigma_c$ baryons.
A rather small effect of the shift down to small $p_{T}$'s has been found. 
The effect for higher excitation 
(spin 3/2 $\Sigma_c$ baryons) is larger than for lower excitations 
(spin 1/2 $\Sigma_c$ baryons) which has purely kinematical origin and is related to masses of the $\Sigma_c$ baryons.
A study of the production (and feed-down) of 
$\Sigma_c$ baryons will be possible with larger statistics by 
the ALICE collaboration \cite{Grelli}.

The independent parton fragmentation approach is only a simplification
which has no firm and fundamental grounds and requires tests 
to be valid approach.
At low energies an asymmetry in production of $\Lambda_c^+$ and 
$\Lambda_c^-$ was observed \cite{Lambdac_asymmetry}.
This may be related to the charm meson cloud in the nucleon \cite{Cazaroto:2013wy}
and/or recombination with proton remnants \cite{recombination}.
At high-energy this mechanism is active at large $x_F$ (or $\eta$,
probably for pseudorapidities larger than available for LHCb.
Certainly a study of $\Lambda_c^+ - \Lambda_c^-$ asymmetry in run II
would be a valueable supplement. This would allow to verify the 
$c \to \Lambda_c$
``independent'' parton hadronization picture.
The new data of the ALICE Collaboration suggests a much bigger
$f_{c \to \Lambda_c}$ hadronization fraction than those obtained
in other processes and LHCb.
In principle, it could be even a creation of $\Lambda_c$ in the
quark-gluon plasma due to coalescence mechanism (see \textit{e.g.} Ref.~\cite{SLS2018}).
Such an enhancement was observed in p-Pb and Pb-Pb collisions and 
interpreted in terms of quark combination/coalescence approach 
in \cite{LSSW2017} (for p Pb) and \cite{PMDG2017} (for Pb Pb).
Even in the "independent" parton picture the hadronization fractions
$f_{c \to D_i}$ or $f_{c \to \Lambda_c}$ do not need to be universal and
may depend on partonic sourounding associated with the collision 
which may be, in principle, reaction and energy dependent. 
Therefore precise measurements at the LHC will allow to verify 
the picture and better understand the hadronization mechanism.

To explore experimentally the hypothesis that $\Lambda_c$ is produced in the mini quark-gluon plasma
one could study its production rates as a function of event multiplicity and compare to
similar analysis for the production of $D^{0}$ mesons. 

\vspace{1cm}

{\bf Acknowledgments}

We are indebted to Alessandro Grelli and Andre Mishke 
for a discussion of the details of the ALICE experimental results 
for $\Lambda_c$ production in proton-proton collisions.
This study was partially supported by the Polish National Science Center
grant DEC-2014/15/B/ST2/02528 and by the Center for Innovation and
Transfer of Natural Sciences and Engineering Knowledge in
Rzesz{\'o}w.


\end{document}